\documentclass[showpacs,aps,prb,floats,twocolumn]{revtex4}
\usepackage[dvips]{graphicx}
\usepackage{amsmath}
\usepackage[dvips]{epsfig}

\begin{document}

\title{Model of the Interplay of Band J-T Effect with Magnetic Order Mediated 
       by Exchange Interaction}
       
\author{G. Gangadhar Reddy$^{a}$, A. Ramakanth$^{a}$, S.K. Ghatak$^{b}$,
        S.N. Behera$^{c}$, \\W. Nolting$^{d}$ and T. Venkatappa Rao$^{a}$}
	
\affiliation{
 $^{a}$Department of Physics, Kakatiya University, Warangal-506009, India\\
 $^{b}$Department of Physics, IIT Kharagpur,  Kharagpur-721302, India\\
 $^{c}$Physics Enclave, HIG-23/1, Housing Board Phase 1, Chandrasekharpur,
  Bhubaneswar-751005, India\\
 $^{d}$Institut f{\"u}r Physik, Humboldt-Universit{\"a}t zu Berlin,12489
       Berlin, Germany}

\date{\today}

\begin{abstract}

\vspace{0.4cm}

A model calculation is presented with the aim to study the interplay between 
magnetic and structural transitions. The model consists of an orbitally
doubly degenerate conduction band and a periodic array of local moments. The
band electrons interact with the local spins via the s-f interaction. The
interaction of the band electrons with phonons is introduced by including band
Jahn-Teller (J-T) interaction.  The model Hamiltonian, including the above 
terms, is solved for the single particle Greens function. In doing
this an ansatz for selfenergy of electrons, which was developed earlier has been
utilized. The quasiparticle density of states (QDOS) and hence the orbital
populations are calculated treating the ferromagnetism of local moments in the mean
field approximation. The critical value of electron-phonon interaction ($G$) for
the appearance of the band J-T distortion is higher in the ferromagnetic state. The
strain appears at a critical temperature ($T_s$) when $G$ is greater than the 
critical value. The onset of ferromagnetism at $T_C$ ($<T_s$) arrests the growth
of the strain. It is concluded that the magnetization hinders the structural 
transition. The quasiparticle density of states are presented to interpret these
results.

\end{abstract}


\maketitle

\newcommand{\be}{\begin{equation}}
\newcommand{\ee}{\end{equation}}

\section{Introduction}

Interplay of order parameters originating from different degrees of freedom of
electrons is very relevant in condensed matter.  One of the recent interests in this
area is related to the mutual influence of band Jahn-Teller (J-T) effect
and ferromagnetic order.  The former results from the removal of orbital
degeneracy and the latter from the lifting of spin degeneracy. The magnetic
order can originate either from the same electrons that are involved in the 
band J-T effect or from the electrons in some other states.  The first case
corresponds to narrow band solids which are usually described by the  
two-fold degenerate Hubbard
model and is applicable to intermetallic compounds.  The interplay
between the above two order parameters had been considered earlier and it was
shown that the magnetic order tends to suppress the spontaneous distortion
resulting from charge-lattice interaction. In the present work, we are
considering a situation where the electrons involved in J-T splitting and the
magnetic order are different.  The doped manganites$^{1,2}$ and Heusler
alloys$^{3,4,5}$ are some of the real systems that represent this case. Doped
manganites shows a rich variety of phenomena such as several forms of magnetic,
orbital and charge ordering$^{6,7}$
The interesting physics of manganites is due to the dynamics of the d-electrons 
of the $Mn$ ion. In the host material $LaMnO_3$, which is a Mott insulator, there 
are four electrons per $Mn$ ion. The d-orbitals are split by the crystal field 
in to two, namely the triply degenerate (ignoring the spin) $t_{2g}$ states 
which are well localized and the doubly degenerate $e_g$ states which are 
extended and form a doubly degenerate band. Out of the four d-electrons, three 
occupy the $t_{2g}$-states and the remaining electron is in the $e_g$-band. 
The $Mn$ ion is in $Mn^{3+}$ state. Due to the large Hund's rule coupling in 
this system, the three $t_{2g}$-electrons have their spins oriented parallel 
to each other making a localized spin of $S=3/2$. This is again strongly Hund's 
rule coupled to the $e_g$-electron. Since the $e_g$ is an extended state, the
electron in this state can hop from lattice site to lattice site. The hopping
combined with Hund's rule coupling is responsible for the long range magnetic
order that exists in this system. This is known in literature as the double
exchange mechanism. Another way of looking at the system is that there is a
localized spin at each lattice site and the band electrons interact with these spins
via an intraatomic exchange interaction. This is known in literature as the
Kondo-lattice model (KLM). 
Similar scenario also prevails in Heusler alloy like $Ni_2MnGa$, where $Mn$
posses localized moment and  the magnetic properties of the host system can be
understood from the KLM with the carrier concentration ($e_g$-electrons) of one
per $Mn$ atom. When the trivalent rare-earth ion is partly replaced by  divalent
ions like $Ca$, $Sr$ or $Ba$, the material exibits CMR properties undergoing transition
from the paramagnetic-insulator state to the ferromagnetic-metal one. The localized spin
of $S=3/2$ is retained but there is a decrease in the electron concentration in
the $e_g$-band. The carrier concentration is less than one per atom. While studying the
CMR and the associated magnetic and insulator-metal transitions, it was realized
that KLM alone is not sufficient to understand the physics of manganites. It is
now accepted that electron-phonon interaction plays an important 
role$^{8,9,10,11}$. 
These materials crystallize in perovskite structure and the Mn ion has $O_6$
octahedron as its immediate environment and therefore interacts strongly with
the distortions in the octahedron. Obviously, this introduces J-T
effect which lifts the degeneracy of the $e_g$-band. The spontaneous distortion
associated with the J-T effect exists when the lowering of the band energy is
more than the increase in elastic energy due to the strain. The simplest way to
describe the band J-T effect is to incorporate the interaction of the
$e_g$-electrons with the lattice distortions(phonons) in the KLM model.  When this
is done, it is pertinent to examine the interplay between the band J-T effect
and magnetism in the model system.  There are experimental results supporting
such an interplay in manganites and in Heusler alloys. It was observed that the J-T
distortion in $(La-Y)CaMnO_3$$^{12}$ and $(La, Pr, Ca)MnO_3$$^{13}$ is reduced
in ferromagnetic state. The suppression of the J-T distortion in
$(Nd,Sm)_{1/2}Sr_{1/2}MnO_3$ in ferromagnetic state under high magnetic field
has also been reported$^{14}$.  The coexistance of the J-T distorted phase and
the ferromagnetic phase have been reported in $Rh_2CoSn$$^{3}$.  J-T splitting exists
in these systems below the transition temperature $T_C$.

Therefore, in order to study the interplay of structural and magnetic transitions,
we consider a model where the band electrons, which are approximated to be s-electrons
in doubly degenerate extended states interact
intraatomically with a periodic array of localized spins (a Kondo
lattice). We provide for the spontaneous lifting of the degeneracy of the band
states by including a band J-T interaction. The lifting of the degeneracy is
signalled by the appearance of strain. The presence of the long range magnetic order
is characterized by the nonzero value of the magnetization.
We study, selfconsistently, the 
strain as a function of the J-T coupling constant for different carrier
($e_g$-electron) concentrations and the dependence of the strain on temperature. In
the latter case, the magnetization of the local moments which is caused by the
exchange interaction between the $e_g$- and $t_{2g}$-electrons (Kondo
interaction) determines, decissively, the temperature dependence of the strain.

\section{Model Hamiltonian and its approximate solution}

The $e_g$-electrons moving in the doubly degenerate band are described by
\begin{eqnarray}
H_{s}&=&\sum_{\alpha=1}^2\sum_{ij\sigma}\left(T_{ij}-\mu\delta_{ij}
\right)c_{\alpha i\sigma}^{\dagger}c_{\alpha j\sigma} \nonumber\\
&=&\sum_{\alpha\mathbf{k}\sigma}(\epsilon(\mathbf{k})-\mu)
c_{\alpha \mathbf{k}\sigma}^{\dagger}c_{\alpha \mathbf{k}\sigma} \qquad
\end{eqnarray}
$T_{ij}$ is the hopping integral for hopping of the electrons from lattice site
$i$ to $j$. $c_{\alpha i\sigma}^{\dagger}$($c_{\alpha i\sigma}$) is the 
creation(annihilation) operator for an electron in the $\alpha$-state on the 
lattice site $i$ with spin $\sigma$. $\alpha=1,2$ is the band index. $\mu$ 
is the chemical potential. $\epsilon(\mathbf{k})$ is the band energy related 
to $T_{ij}$ by
\be
T_{ij}=\frac{1}{N}\sum_\mathbf{k}\epsilon(\mathbf{k})e^{i\mathbf{k}
\cdot(\mathbf{R}_i-\mathbf{R}_j)}
\ee
The band electrons interact with the localized spins via the intraatomic
exchange interaction of the coupling strength $J$ and this is described by
\begin{eqnarray}
H_{sf}&=&-J\sum_{j,\alpha}\mathbf{\sigma}_{\alpha,j}\cdot \mathbf{S}_{j}
\nonumber\\
&=&-\frac{1}{2}J\sum_
{\alpha,j\sigma}(z_{\sigma}S_{j}^{z}n_{\alpha j\sigma}+S_{j}^{-\sigma}
c_{\alpha j-\sigma}^{\dagger}
c_{\alpha j\sigma})
\end{eqnarray}
$\mathbf{\sigma}$ is the spin of the band electron and $\mathbf{S}$ is the localized 
spin (total spin of the three $t_{2g}$ electrons). $n_{\alpha j\sigma}$ is the
number operator for the electron in the state $\alpha$ at the lattice site $j$
with spin $\sigma$. At the outset itself we assume a ferromagnetic interation
($J>0$). $z_\sigma$ is a sign factor, $z_{\sigma}=\delta_{\sigma
\uparrow}-\delta_{\sigma\downarrow}$ 
and $S_{j}^{\sigma}=S_{j}^{x}+iz_
{\sigma}S_{j}^{y}$.

The electron density in the degenerate band couples to the static elastic
strain through the J-T interaction. In the case of a tetragonal distortion, 
this interaction is described by$^{15,16,17}$
\be
H_{JT}=Ge\sum_{\mathbf{k},\sigma}(n_{1\mathbf{k}\sigma}-n_{2\mathbf{k}\sigma})
=Ge\sum_{i\sigma}(n_{1i\sigma}-n_{2i\sigma}).
\ee
$G$ is the strength of the J-T coupling and $e$ is the lattice strain given by
\be
e=\frac{G}{NC_0}\sum_{i\sigma}(<n_{1i\sigma}>-<n_{2i\sigma}>)
\ee
where $C_0$ is the elastic constant. It is clear that $H_{JT}$ tries to create a difference in the occupation of the
two degenerate bands. The difference in occupation leads to the building up of
the strain. Thus, under suitable conditions, there is a spontaneous splitting of
the bands and building up of strain which indicates a structural transition. The
building up of the strain however leads to an increase in the lattice elastic
energy which is given by
\be
H_L=\frac{1}{2}NC_0e^2
\ee
Where $N$ is the total number of atoms. Since
this term is a c-number and we are not looking for the ground state whose
energy has to be minimum, we leave this term out of our consideration.
Then the Hamiltonian of the model system we are considering is
\be
H=H_{s}+H_{sf}+H_{JT}.
\ee 
The model Hamiltonian Eq(7) obviously cannot be solved exactly. However, in an
earlier work$^{18}$, we have proposed, for the Hamiltonian without the J-T term, 
an approximation scheme, which is reliable in the limit of low carrier 
concentration. We will exploit that scheme in solving the present model. 
Firstly, without resorting to any approximation, we can absorb the J-T term 
into $H_s$ by modifying the band energies for the two bands as
\be
\epsilon_\alpha(\mathbf{k})=\epsilon(\mathbf{k})+(-1)^\alpha Ge.
\ee
Then we have 
\be
\mathcal{H}_{s}=\sum_{\alpha\mathbf{k}\sigma}(\epsilon_\alpha(\mathbf{k})-\mu)
c_{\alpha \mathbf{k}\sigma}^{\dagger}c_{\alpha \mathbf{k}\sigma}
\ee
and the total Hamiltonian is given by
\be
H = \mathcal{H}_s+H_{sf}.
\ee
In order to calculate the strain ( caused by the structural transition), 
one has to calculate the one-electron Greens function
\begin{eqnarray}
G_{\alpha\mathbf{k}\sigma}(E)&=&<<c_{\alpha \mathbf{k}\sigma}:
c_{\alpha \mathbf{k}\sigma}^{\dagger}>>_E \nonumber\\
&=& 
\frac{1}{E-\epsilon_\alpha(\mathbf{k})-\Sigma_{\alpha\sigma}(E)}.
\end{eqnarray}
That means, one has to calculate the selfenergy $\Sigma_{\alpha\sigma}(E)$ of the 
electron in the presence of $H_{sf}$. There are some exact results available for 
the selfenergy in certain limiting cases, namely, the zero band width limit for all
temperatures$^{19}$ and the finite band width but ferromagnetic saturation ($T=0$) 
limit$^{20,21}$. In addition, using Mori formalism, the result for the second order 
perturbation theory is also available$^{22}$. We  propose an ansatz for the self 
energy that reproduces the known limiting results  and in addition,  satisfies
the strong coupling limit. This can be taken care  of by making a high energy
expansion and evaluating the first four spectral moments$^{22}$. We thus have a
selfenergy which fulfills a) the zero band width limit for all temperatures and
coupling strengths, b) $T=0$ limit for all band widths and coupling constants,
c) the weak coupling limit ($\mathcal O J^2$) and  d)has the correct high 
energy behaviour. The
ansatz for $\Sigma_{\alpha\sigma}(E)$ is given by  
\be
\Sigma_{\alpha\sigma}(E)=-\frac{1}{2}Jm_\sigma+\frac{1}{4}J^2\frac
{a_\sigma G_{\alpha 0}\left(E-\frac{1}{2}Jm_\sigma\right)}
{1-b_\sigma G_{\alpha 0}\left(E-\frac{1}{2}Jm_\sigma\right)}.
\ee 
Here
\be
G_{\alpha 0}(E)=\frac{1}{N}\sum_\mathbf{k}G_{\alpha\mathbf{k}}(E)=
\frac{1}{N}\sum_\mathbf{k}\frac{1}{E-\epsilon_\alpha(\mathbf{k})}
\ee
$m_\sigma=z_\sigma\left<S^z\right>$. The ansatz assumes a
$\mathbf{k}-$independent selfenergy. 
As $H_{sf}$ is a local interaction, the energy dependence of the selfenergy is
the deciding factor in relation to the electron density of states. 
The parameters $a_\sigma$ and $b_\sigma$ and are fixed by rigorous high
energy expansions to fulfill the first four spectral moments:
\be
a_\sigma=S(S+1)-m_\sigma(m_\sigma+1)\hspace{0.7cm}b_\sigma=b_{-\sigma}
=\frac{1}{2}J
\ee
It should be mentioned that this ansatz is valid only in the limit of
low carrier density. Since we are interseted in simulating systems with
low carrier density, it is justified to use the above ansatz.
From $G_{\alpha\mathbf{k}\sigma}(E)$ one can obtain the spectral density 
$S_{\alpha\mathbf{k}\sigma}(E)$ and the density of states 
$\rho_{\alpha\sigma}(E)$ from the well known relations
\begin{eqnarray}
S_{\alpha\mathbf{k}\sigma}(E)&=&-\frac{1}{\pi}Im G_{\alpha\mathbf{k}\sigma}
(E)\\
\rho_{\alpha\sigma}(E)&=&\frac{1}{N}\sum_\mathbf{k}
S_{\alpha\mathbf{k}\sigma}(E)
\end{eqnarray}

\noindent From the knowledge of the density of states, the expectation
values can be evaluated:
\be
\left<n_{\alpha\sigma}\right>=\int dEf_{-}(E)\rho_{\alpha\sigma}(E)
\ee

\noindent  Where $f_{-}(E)=1/\left(1+e^{\beta E}\right)$ is the Fermi function with
 $\beta=1/kT$.
The chemical potential $\mu$ is fixed by the constraint
 \be
 n=\sum_{\alpha\sigma}\left<n_{\alpha\sigma}\right>=constant.
 \ee
 
\noindent For a given set of the model parameters $n$, $J$, $S$ and $G$ and
for a fixed $T$, the occupencies of both lower and upper
sub-bands for each spin directions is computed self-consistently. After having
the self consistent solution, the lattice strain 

\be
e =\frac{G}{C_0}\sum_{\sigma}\left(\left<n_{1\sigma}\right>-\left<n_{2\sigma}\right>\right)
\ee

\noindent is calculated. The average occupation of $e_g$-orbitals $\langle
n_{\alpha\sigma}\rangle$ can be
numerically obtained using a model density of states  for the "free" $e_g$-band:
\be
\rho_0(E) = A \sqrt {1-\left|\frac {E}{D}\right|}\
\ln\left|\frac{D^2}{E^2}\right|
\ee
Where $A$ is a normalization constant and $D$ is half the width of free Bloch band.
In order to calculate $\langle n_{\alpha\sigma}\rangle$, we require $\langle
S^z\rangle$, since this enters into the selfenergy. The local moment system is
described within the mean field approximation and is represented by Brillouin
function. The effective field seen by the local moment is determined by mutual
exchange interaction which fixes $T_C$. Numerical results are given where all
the energy parameters are normalized in terms of the free bandwidth ($2D$).  
Ideally, one should get the magnetization $\left<S^z\right>$ selfconsistently
out of the calculation. However, it is a very involved problem. Therefore, we
treat it as a parameter and obtain its value at any temperature from the 
Brillouin function assuming a value for $T_C$.

Using the above theory, the results for the density of states and the strain are
presented in the next section.

\section{Results and discussion}

First we consider the $T=0$ case for two extreme situations of the local
magnetization, namely, the paramagnetic ($\langle S^z\rangle=0$) and the saturated
ferromagnetic ($\langle S^z\rangle=S$) state. In Fig.1 we have plotted the strain as
function of the J-T coupling constant $G$. We find that unless the value of $G$
exceeds a critical value, there is no spontaneous splitting of the orbitally degenerate
bands. The splitting of the  bands takes place and the
lower of the split bands is occupied more by the electrons than the upper
one in order to lower energy.
This is energetically favoured only for a sufficiently large $G$. When $G$ is
further increased, the lower of the split bands is more populated and therefore
the strain increases as shown in the figure.  Though not shown in the
figure, the critical value of $G$ also depends on the carrier concentration. 
Larger is $n$,
smaller is the critical value of $G$. Another feature which is displayed in
\begin{figure}[thbp]
\begin{center}
    \epsfig{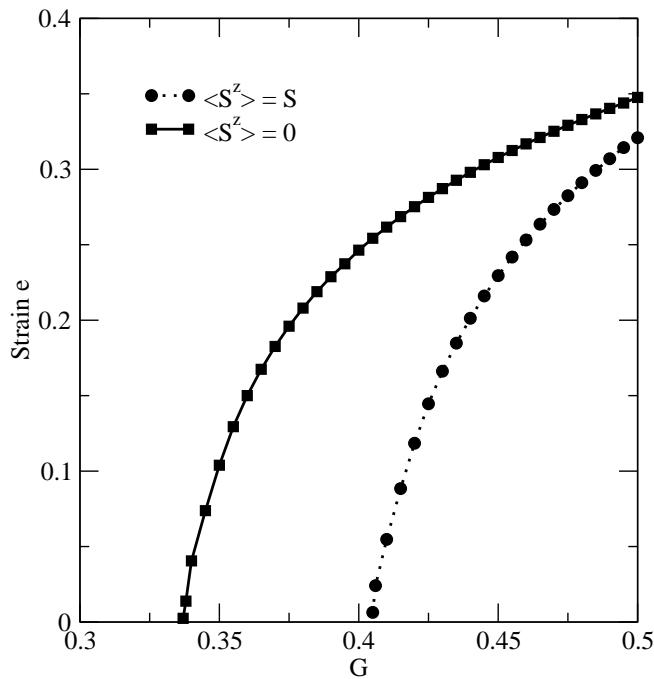}
    \caption{Lattice strain
    as a function of electron-phonon coupling constatnt $G$ at $T = 0$ with and
    without localized magnetization.
    $n = 0.7$, $J = 1$, $S= 3/2$ and  $C_0 = 1$. }    
\end{center}
\end{figure}
Fig.1, is the role of the magnetization $\left<S^z\right>$.
Critical value of $G$ to induce J-T effect is much higher for the saturated
local magnetization situation as compared to that of the paramagnetic situation.
With increase of $G$, it is observed that the gap between the two curves
decreases. The difference in the critical value of $G$ appears to be a result of
the competition between the mechanisms leading to the lifting of orbital and spin
degeneracies of the $e_g$-state. In the ferromagnetic situation, the effective
field lifts the spin degeneracy so that stronger electron lattice interaction
is necessary for the strain to appear. As the strain becomes larger, the energy
gain by the J-T distortion dominates over the energy gain by lifting of the spin
degeneracy. So the influence of magnetization becomes less important. The
interplay becomes more interesting when the energy gain due to the two
mechanisms is comparable.

In order to examine this interplay in more detail, we study the temperature
dependence of the strain in Fig. 2. 
\begin{figure}[thbp]
\begin{center}
    \epsfig{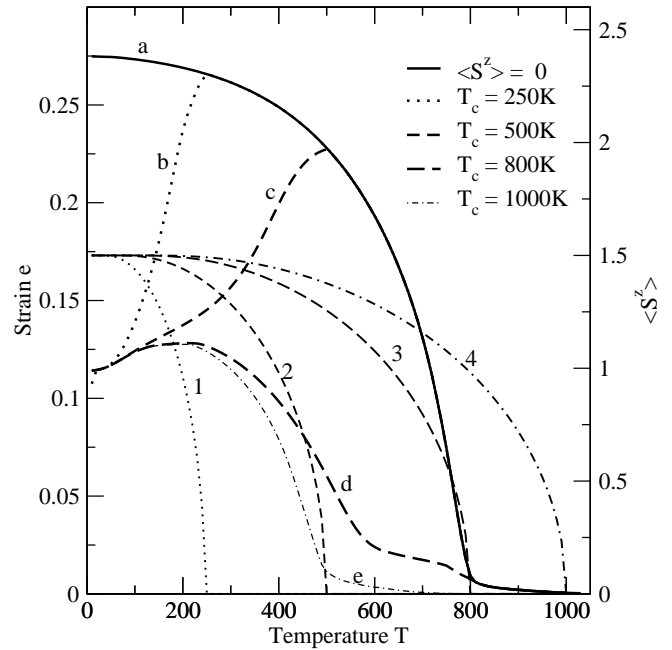}
    \caption{Lattice strain
    as a function of temperature for various values of $T_c$ (curves a, b, c, d
    and e).
    $n = 0.7$, $J = 1$, $S= 3/2$, $C_0 = 1$ and $G = 0.42$. Brillouin function is
    also plotted(curves 1,2,3 and 4) in the same figure as a guide for the eye.}    
\end{center}
\end{figure}
The structural transition temperature $T_s$ is the temperature at which the strain
goes to zero. We choose the parameters such that $T_s=800K$.
Then we study the effect of magnetization
on strain by varying $T_C$ such that i) $T_C<T_s$, ii) $T_C\approx T_s$ and iii)
$T_C>T_s$. The Fig. 2 displays the T-dpendence of the strain(curves a, b c ,d and e)
and the magnetization (which is a Brillouin function for a given $T_C$) for 
different $T_C$'s (curves 1, 2, 3, and 4). Whatever is $T_C$, at $T=0$,
 $\langle S^z\rangle=S$. Therefore, the effect of the magnetization on the strain is
 independent of $T_C$ and leads to a maximum decrease of strain. As $T_C$
 increases, the strain goes up with increasing $T$ and the rate of increase is
 higher for lower $T_C$ so long as $T_C<T_s$. therefore, there always appears a
 peak in the curves. When $T_C\approx T_s$, the peak is very faint and the
 strain becomes very small well before $T_s$ is reached. For $T_C>T_s$, the strain is
 nonzero only when $T$ is much lower than $T_C$ (curve e of Fig.2).
 
 It is clear that at $T=0$, the presence of magnetization causes a
 redistribution of electrons between the orbital levels by creating a population
 difference between the spin levels. Such redistribution is the cause  of
 suppression of strain. As $T$ increases, the spin level occupancies tend to equalize
 and that polarises the orbital levels further resulting in the increase in
 the strain compared to its $T=0$ value. Since for smaller $T_C$ the magnetization
 decreases faster with increasing $T$, the increase in strain is also faster.
 For $T_C>T_s$, due to the choice of parameters, as expected, there is no strain
 between $T_s<T<T_C$. When $T$ is much lower than $T_C$, the occupancy of the spin
 levels is stabilized and the system can lower energy by further redistribution
 of electrons between the orbital states. That is why the strain is finite for $T$ much
 less than $T_C$ (curve $e$ of Fig. 2).
\begin{figure}[thbpb]
\begin{center}
    \epsfig{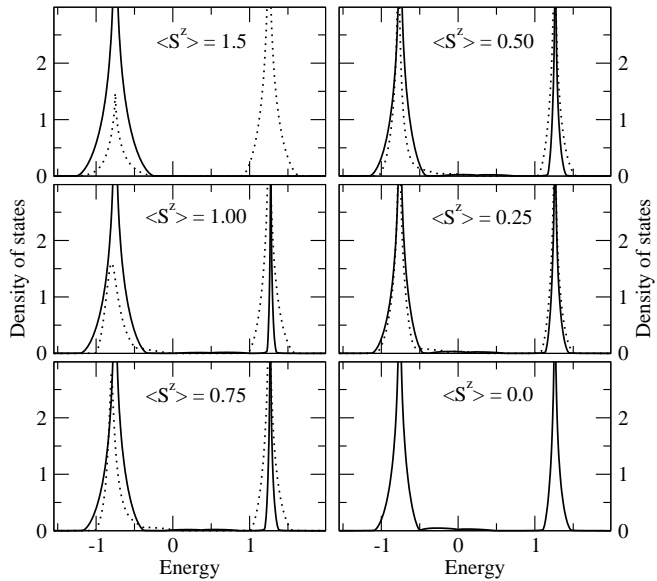}
    \caption{Quasiparticle density of states as a function 
     of enregy for various values of localized magnetization $\langle S^z\rangle$ in the
     absence of Jahn-Teller distortion i.e for $G=0$. Full line for spin
    up and dotted line for spin down.  $J = 1$ and  $S= 3/2$}    
\end{center}
\end{figure}
\begin{figure}[thbp]
\begin{center}
    \epsfig{file=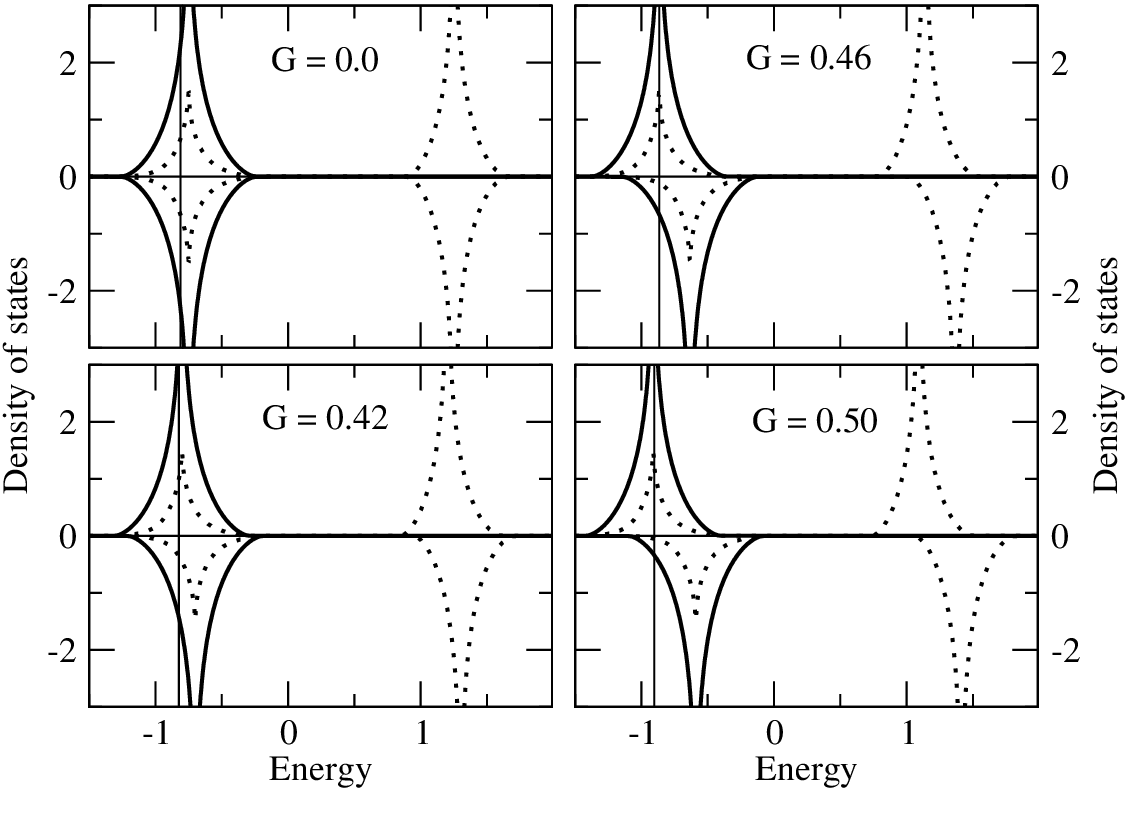, width=0.48\textwidth}
    \caption{Quasiparticle density of states (for lower sub-band($\alpha = 1$) 
     in the positive half of the
     frame and for upper sub-band($\alpha = 2$) in the negative half of the frame) as a function 
     of enregy for various values of electron-phonon coupling constatnt $G$. Full line for spin
    up and dotted line for spin down. Thin vertical line shows the position of
    the chemical potential. $n = 0.7$, $J = 1$, $S= 3/2$, $C_0 = 1$, $\langle
    S^z\rangle = S$ and $T = 0$.}        
\end{center}
\end{figure} 
The results obtained are interpreted with the help of the quasiparticle density of
states (QDOS).
Before discussing the results of the full problem, to fix up a reference for
further discussion, we want to present in Fig.3, the QDOS for the case of a 
pure KLM, that is, for the case of $G=0$. The QDOS consists of two subbands for
each spin direction separated by an energy of the order of $\frac{1}{2}J(2S+1)$.
The separation of the bands is independent of $T$ but the spectral weights of 
these subbands, however, depend on $T$ through the
value of $\left<S^z\right>$. For example, at $T=0$ ($\left<S^z\right>=S$), the
spectral weight of the upper subband for $\uparrow$-states is zero.The reason
for this is easy to understand.  At $T=0$, the local moment system is
saturated.  Therefore, for an $\uparrow$-electron there is no chance to spin-flip
 by involving a corresponding spin-flip of the local moment system.  That
means, at $T=0$, as far as the $\uparrow$-electron is concerned, only the Ising
part of $H_{sf}$ operates resulting simply in a rigid shift of QDOS.  As we see
from Fig.3, the spectral weight of the $\downarrow$-states in the lower sub band
is finite.  This is because, for a $\downarrow$-electron, even at $T=0$,
spin-flip is possible.  Further more, when a $\downarrow$-electron flips its
spin, it lands as an $\uparrow$-electron.  Therefore the nonzero QDOS of the
$\downarrow$-electron should be in the same energy region as that of
$\uparrow$-electron.  For a $\downarrow$-electron there is another possibility. 
It can have repeated magnon emission and absorption.  That is, in a sense, it
propagates in the lattice dressed by a cloud of magnaons.  This is a stable
quasiparticle, which we call the magnetic polaron.  Obviously, at $T=0$, there is 
no possibility of magnetic polaron for $\uparrow$-electron.  As $T$ increases
($\left<S^z\right>$ decreases), the
spin flip processes are allowed for both the spin directions and therefore the
spectral weights in both the subbands are nonzero for both the spin directions. 
At $T=T_C$ ($\left<S^z\right>=0$), the spectral weights of $\uparrow$- and 
$\downarrow$-states in the two subbands become equal as it should be. We note
the assymetry with respect to the centre of the free band. This originates from
the renormalization of the atomic levels by the s-f interaction$^{19}$.

Now we consider the further splitting of these bands due to the J-T effect. That 
is, when the degeneracy of the $e_g$ band is lifted due to the J-T effect, each of 
the subbands of Fig.3 for each spin direction discussed above is again split 
into two as shown in Fig.4.  
Noting the position of the chemical potential, we see that the upper subbands for
both values of $\alpha$
can be ignored since they are never populated. Only we have to keep in mind the
change in their spectral weights. From now on, when we speak of subbands they
are the two $e_g$ bands, corresponding to $\alpha = 1,2$ as displayed in the
positive and negative halves of the frame. 

We want to understand the dependence of the strain on the coupling constant $G$ and
the influence of $\left<S^z\right>$. In Fig.4, we take the case of saturation 
($\left<S^z\right>=S$). When $G$ is less than the critical value, the two $e_g$
bands are degenerate. As $G$ approaches the critical value of 0.42, there is a
slight splitting of the two subbands and the difference in their population becomes
non zero. At the same time, it should be noticed that the position of both the
subbands shifts to lower energy, so that, on the whole, the energy of the system is
lowered by the splitting. For a small increase in $G$, the splitting increases
and the increase in strain is very large.  Further, the shifting of 
the subbands to lower
energy is also large so that the energy of the system is much lower. Any further
increase in $G$ does not have much effect on the quasiparticle spectrum any more
which means the strain saturates. It should be emphasized that the strain is not
introduced by hand but comes selfconsistently out of the model.

\begin{figure}[thbp]
\begin{center}
    \epsfig{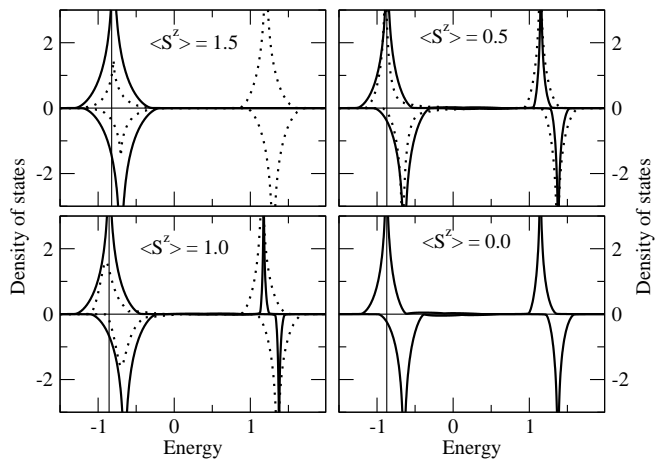}
    \caption{Quasiparticle density of states(for lower sub-band in the positive half of the
     frame and for upper sub-band in the negative half of the frame) as a function 
     of enregy for various values of localized magnetization $\langle S^z\rangle$. Full line for spin
    up and dotted line for spin down. Thin vertical line shows the position of
    the chemical potential. $n = 0.7$, $J = 1$, $S= 3/2$, $C_0 = 1$, $T = 0$ and
    $G = 0.42$.}
\end{center}
\end{figure}

\begin{figure}[thbp]
\begin{center}
    \epsfig{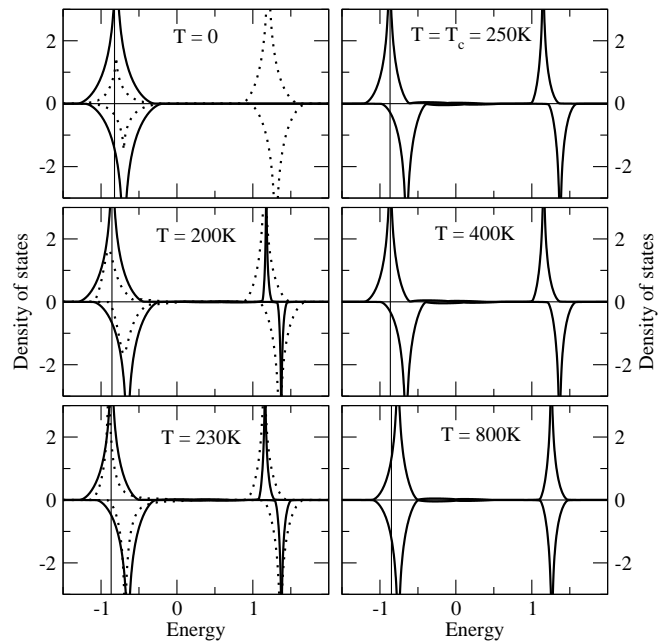}
    \caption{Quasiparticle density of states(for lower sub-band in the positive half of the
     frame and for upper sub-band in the negative half of the frame) as a function 
     of enregy for various values of temperature $T$. Full line for spin
    up and dotted line for spin down. Thin vertical line shows the position of
    the chemical potential. $n = 0.7$, $J = 1$, $S= 3/2$, $C_0 = 1$, and $G = 0.42$ .}    
\end{center}
\end{figure}

In Fig.5 we now fix the value of $G$ at 0.42, so that the model prefers the J-T split
situation and look at the influence of $\left<S^z\right>$. 
At saturation ($\left<S^z\right>=S$), the spectral weight of $\uparrow$-subband
is such that the $\alpha=1$ subband is slightly more populated compared to
the $\alpha=2$ subband. As a result, a small strain appears. As 
$\left<S^z\right>$ decreases, this spectral weight is modified in such a way
that the occupation of the $\alpha=1$ subband is more than that of the $\alpha=2$. This
results in an increase of the strain with $T$ as depicted in Fig. 2.  This trend
 continues until 
$\left<S^z\right>=0$ and at that point, the strain is as if there is no exchange
interaction in the model. In short, the effect of magnetization is to hinder J-T
splitting.

In order to examine the interplay of magnetization and the strain as a function
of $T$, it is necessary to include the temperature variation of both the strain and
the magnetization. Therefore, we consider a specific case of $T_C=250K$ and
$T_s=800K$ and the corresponding QDOS at different temperature are displayed in
Fig. 6. Starting from $T=0$, the $\alpha=1$ subband occupation increases as $T$
increases upto $T_C$. Above $T_C$, the situation is reversed and  both the 
$\alpha$-subbands are equally occupied at $T_s$. Therefore, the maximum of
strain occurs at $T_C$ as shown in curve (b) of Fig. 2.

\section{Conclusions}

The model Hamiltonian is solved by first absorbing the J-T term into the band
electron term and then utilizing an interpolation ansatz for the selfenergy. The
band splitting and through it the strain in the lattice due to J-T interaction
has been determined selfconsistently. The strain as a function of the coupling
constant at $T=0$ is studied with and without the presence of local moment ordering. It
is found that a minimum value of $G$ is required for the strain to appear. This critical G
is larger if the local moments are ordered or alternately if there is an
external magnetic field. The temperature dependence of the strain is studied by
assuming different values for the magnetic transition temperature. The study
indicates that there is a strong interplay between the magnetic and structural
transitions. It is observed that the growth of the strain appearing at $T_s$
($>T_C$) is arrested with the onset of ferromagnetism and tends to a lower value
determined by the magnetization at $T=0$. This means the removal of spin
degeneracy is not conducive to the removal of orbital degeneracy. The results
are explained on the basis of the QDOS. The basic ingredients of the model are 
the band J-T effect, long range magnetic order and their mutual interaction. 
The possibility of different hopping between the degenerate bands is not
considered for simplicity and therefore the results correspond to large
J-T effect.  The inter-orbital hopping, which is sometimes considered, would
remove partially the
degeneracy of the state.  Also the magnetism which should evolve within the
model is treated as a parameter. The detailed comparison with experiment is
therefore not attempted.  However, the general trends of the results
related to suppression of J-T strain and coexistance of two phases are in tune
with the experimental$^{12,13,14}$ observations in a $Ca$-doped manganite and Heusler alloy. 
The calculation is based on the ansatz used for selfenergy which is  valid only
in the limit of low charge carrier concentration. If this needs to be relaxed,
the interaction among the band electrons has to be taken into account and 
naturally the ansatz for the selfenergy has to be modified. 

\section{Acknowledgements}

The authors (GGR, ARK and SKG) are grateful to the Council of Scientific and
Industrial Research, New Delhi, India for financial support 
through grant (No.03(1068)/06/EMR-II) of a scheme

\section{References}

\noindent
$^1$S. Jin, T.H. Tiefel, M. Mc Cormack, R.A. Fastnacht, R. Ramesh and L.H. Chen
 Science \textbf{264}, 413 (1994)

\noindent 
$^2$A.P. Ramirez, J. Phys.: Condens. Matter \textbf{9}, 8171 (1997)

\noindent
$^{3}$J.C. Suits, Solid State Commun. \textbf{18}, 423 (1976). 

\noindent
$^{4}$ Shinpei Fhjii, Shoji Ishida and Setsuro Asano, J. Phys. Soc. of Japan
\textbf{58}, 3657 (1989).

\noindent
$^{5}$K. Ooiwa, K. Endo and A. Shinogi, J. Mag. Mag. Mat. \textbf{104-107}, 2011
(1992).

\noindent
$^6$\textit{Colossal Magnetoresistance Oxides}, edited by Y. Tokura 
(Gordon and Breach, New york, 2000)

\noindent
$^7$M.B. Salamon and M. Jaime, Rev. Mod. Phys. \textbf{73}, 583, (2001)

\noindent
$^8$A.J. Millis, Pb. Littlewood and B.I. Shraiman, Phys. Rev. Lett. \textbf{74},
3164 (1995)

\noindent
$^9$A.J. Millis, B.I. Shraiman and R. Mulller, Phys. Rev. Lett., \textbf{77}, 175
(1996)

\noindent
$^{10}$A.J. Millis, R. Muller and B.I. Shraiman, Phys. Rev. B\textbf{54}, 5389
(1996)

\noindent
$^{11}$A.J. Millis, R. Muller and B.I. Shraiman, Phys. Rev. B\textbf{54}, 5405
(1996)

\noindent
$^{12}$ J.L. Garcia-Munoz, M. Suaaidi, J. Fontcuberta and J. Rodriguez-Carvajai
Phys. Rev.  B\textbf{55}, 34 (1997).

\noindent
$^{13}$ H.J. Lee, K.H. Kim, M.W. Kim, T.W. Noh, B.G. Kim, T.Y. Koo,S.-W. Cheong,
Y.J. Wang and X. Wei, Phys. Rev.  B\textbf{55}, 34 (1997).

\noindent
$^{14}$Y. Tokura, H. Kuwahara, Y. Moritomo, Y. Tomioka and A. Asamistu, 
Phys. Rev. Lett. \textbf{76}, 3184 (1996).

\noindent
$^{15}$D.K. Ray and S.K. Ghatak, Phys. Rev., B\textbf{36}, 3868 (1987)

\noindent
$^{16}$H. Ghosh, M. Mitra, S.N. Behera and S.K. Ghatak, Phys. Rev. B\textbf{57},
13414 (1998)

\noindent
$^{17}$N. Parhi, G.C. Rout and S.N. Behera to appear in Int. J. Mod. Phys B
(2006)

\noindent
$^{18}$W. Nolting, G.G. Reddy, A. Ramakanth and D. Meyer, Phys. Rev. B\textbf{64},
155109 (2001)

\noindent
$^{19}$W. Nolting and M. Matlak, Phys. Status Solidi B\textbf{123}, 155 (1984)

\noindent
$^{20}$B.S. Shastry and D.C. Mattis, Phys. Rev. B\textbf{24}, 5340 (1981)

\noindent
$^{21}$W. Nolting, U. Dubil and M. Matlak, J. Phys. C \textbf{18}, 3687 (1985)

\noindent
$^{22}$G. Bulk and R.J. Jelitto, Phys. Rev. B\textbf{41}, 413 (1990)

\end{document}